\documentstyle[preprint,prb,aps]{revtex}
\tightenlines

\title{Classical and Quantum Behavior of Generalized Oscillators \\ - action variable, angle variable and quantum phase -}

\author{M. Omote$^{1)}$, S. Sakoda$^{2)}$ and S. Kamefuchi$^{3}$\\
$^{1}$ Department of Physics, Keio University, Hiyoshi, Yokohama\\
$^{2}$ Department of Mathematics and Physics, National Defense Academy,
Yokosuka\\
$^{3}$ Atomic Energy Research Institute, Nihon University, Kanda-Surugadai, Tokyo}

\begin{document}

\maketitle
\begin{abstract}
The relation that exists in quantum mechanics among action variables, angle variables and the phases of quantum states is clarified, by referring to the system of a generalized oscillator. As a by-product, quantum-mechanical meaning of the classical Hamilton-Jacobi equation and related matters is clarified, where a new picture  of quantum mechanics is introduced, to be called the Hamilton-Jacobi picture.
\end{abstract}

  
\section{Introduction}

In the correspondence-theoretical arguments of the early quantum theory the so-called action and angle variables played  a very important role \cite{HG}: the quantum condition used to be imposed on the former variables, owing to their property  of being adiabatic invariants. And the purpose of the present  paper is then to study the quantum-mechanical aspects thereof, that is, the problem as to where and how these variables come into play in the conventional formalism of quantum mechanics. As will be seen in what follows, the behavior of these variables are closely related to  phases of quantum mechanical states.  

Recently, adiabatic invariants have further been utilized in the study of  some related problems , such as the Berry phase \cite{MVB}, the Hannay angle  \cite{JHH} etc. . In this connection, we should note, however, that adiabatic invariants are merely approximate ones, being possible only in the so-called adiabatic approximation. Thus, in the following discussions we shall employ, instead of such quantities, something similar but strictly conserved. 

In carrying out such a program  we shall be exclusively concerned  with a relatively simple system of what we shall call a $generalized~ oscillator$, since a discussion of very general systems does not seem to be an easy matter. Now, by a generalized oscillator we mean the system having a classical Hamiltonian such as 
\begin{equation}
H(t)=X(t)p^{2}+2Y(t)pq+Z(t)q^{2},
\end{equation}
where $p$ and $q$ are the usual canonical variables, and $X(t), Y(t)$ and $Z(t)$ are arbitrary real functions of time $t$. That is to say, our system has, in general, a time-dependent Hamiltonian, so that its energy is not necessarily a conserved quantity.

In Sect.2 we shall discuss the classical mechanics of the system, and consider two cases of (generalized) action variables which are given as  a bilinear or linear form of $p$ and $q$. In obtaining formal solutions to the equation of motion we find it convenient to use the method of canonical transformations. In our discussions a quantity $A(t)$ which satisfies a nonlinear differential equation plays a  basic role. The angle variables for  general cases are no longer  linear functions in $t$.

In Sect. 3 we go over to the quantum mechanics  of the system. The method we developed before for systems with time-dependent Hamiltonians are very useful here \cite{OK}. In the expression for formal solutions to the Schr$\ddot{{\rm o}}$dinger equation the angle variable enters into the phase of the corresponding states. Some properties of coherent states are discussed in Sect. 4. There are many ways of constructing  these, each having the right classical limit.

Corresponding to  a classical canonical transformation whose generating function satisfies the Hamilton-Jacobi equation, we go, in Sect. 5, from the Schr$\ddot{{\rm o}}$dinger picture to a new one, to be called the Hamilton-Jacobi picture,   where the transformed quantum Hamiltonian vanishes.  Sect. 6 is devoted to some applications, and Sect. 7 to final remarks.

\section{Classical Solutions to the Equation of Motion}
The equation of motion of this system is given by
\begin{eqnarray}
&& \ddot{q}-\frac{\dot{X}}{X}\dot{q}+\Omega q=0, \\
&&  ~~ \Omega=\Omega(t) \equiv  \frac{2}{X}(\dot{X}Y-X\dot{Y})+4(XZ-Y^{2}) .\nonumber
\end{eqnarray}
In order to find formal solutions to this equation the method of canonical transformations is useful. What we have to do with this method is to find a solution $S$ of the Hamilton-Jacobi equation
\begin{equation}
X~\Bigl(\frac{\partial S}{\partial q}\Bigr)^{2}+2Y q \frac{\partial S}{\partial q}+Z q^{2}+\frac{\partial S}{\partial t}=0,
\end{equation}
where $S=S(q,t)$, Hamilton's principal function, is a function of $q$ and $t$.

In the case of the Hamiltonian $H$ being a constant of motion, the Hamilton-Jacobi equation (3) can be solved by assuming  $S(q,t) \equiv \tilde{S}(q)-E t$, where $\tilde{S}(q)$ is Hamilton's characteristic function and $E$ a constant \cite{HG,VIA}. The solutions of the equation of motion can be obtained by  the canonical transformation generated by $S(q,t)=\tilde{S}(q)-Et$. 

On the other hand, the above method cannot be used for those cases in which the Hamiltonian $H(t)$  depends on $t$ explicitly. In order to find a way out,  we consider, in the following,  two kinds of canonical transformations which prove to be useful for finding   formal solutions of (2).

\subsection{Canonical transformation 1}

We begin by introducing a quantity $P$ which is defined as a bilinear form of $p$ and $q$  
\begin{equation}
P=A(t) p^{2}+2B(t) pq +C(t)q^{2}
\end{equation}
where coefficients $A(t) (>0), B(t)$ and $C(t)$ are some suitable  real functions of the time variable $t$, and then  consider a canonical transformation $(p,q) \to (P,Q)$. The generating function $S_{1}(q,P,t)$ of the canonical transformation is a function of $q,P$ and $t$, and canonical variables $p$ and $Q$ are given by  
\begin{equation}
p=\frac{\partial S_{1}(q,P,t)}{\partial q},~~~Q=\frac{\partial S_{1}(q,P,t)}{\partial P}.
\end{equation}
From (4) and (5) we have a differential equation for $S_{1}(q,P,t)$:
\begin{equation}
\frac{\partial S_{1}(q,P,t)}{\partial q}=-\frac{B(t)}{A(t)}q+\frac{1}{A(t)}\sqrt{A(t)P-\kappa q^{2}},
\end{equation}
where $\kappa$ is defined by $\kappa(t) \equiv A(t)C(t)-B^{2}(t)$.

For $\kappa$ there are three cases :  $\kappa >0$,  $\kappa=0$ and  $\kappa <0$ . In the following  we consider the case  $\kappa >0$ only, and then    find  by integrating (6)
\begin{equation}
S_{1}(q,P,t)=-\frac{B  }{2A }q^{2}+ \frac{\sqrt{\kappa}}{2 A }\Bigl[q\sqrt{\frac{A P}{\kappa}-q^{2}}+\frac{A P}{\kappa}\sin^{-1}\Bigl(\sqrt{\frac{\kappa}{A P}}q\Bigr)\Bigr]+s_{1}(P,t),
\end{equation}
where $s_{1}(P,t)$ is an arbitrary function of $P$ and $t$. From (5) and (7) the canonical coordinate  variable $Q$, conjugate to $P$, is given by
\begin{equation}
Q=\frac{1}{2\sqrt{\kappa}}\sin^{-1}\Bigl(\sqrt{\frac{\kappa}{A P}}q\Bigr)+\frac{\partial s_{1}(P,t)}{\partial P} .
\end{equation} 
Then the canonical variables $q$ and $p$ are found,  in terms of $Q$ and $P$, to be
\begin{eqnarray}
q&=&\sqrt{\frac{AP}{\kappa}} \sin \Bigl\{2\sqrt{\kappa}\Bigl(Q-\frac{\partial s_{1}(P,t)}{\partial P}\Bigr)\Bigr\},  \\
p&=&\sqrt{\frac{P}{A}}\Bigl\{-\frac{B}{\sqrt{\kappa}}\sin \{2\sqrt{\kappa}(Q-\frac{\partial s_{1}(P,t)}{\partial P})\}+\cos \{2\sqrt{\kappa}(Q-\frac{\partial s_{1}(P,t)}{\partial P})\} \Bigr\}.
\end{eqnarray} 
The transformed Hamiltonian $K_{1}=K_{1}(Q,P,t)$ is given by
\begin{equation}
K_{1}(Q,P,t)=H(t)+\frac{\partial S_{1}}{\partial t} .
\end{equation}
By substituting (9), (10) and (7) into (11) we have 
\begin{eqnarray}
K_{1}(Q,P,t)&=&\frac{P}{A}X +\frac{\partial s_{1}(P,t)}{\partial t}  \nonumber  \\
&&+\frac{P}{2\kappa A}\Bigl\{2B^{2}X -4ABY +(B\dot{A}-A\dot{B})+2A^{2}Z -2\kappa X \Bigr\}\sin^{2} \Bigl\{2\sqrt{\kappa}\Bigl(Q-\frac{\partial s_{1}(P,t)}{\partial P}\Bigr)\Bigr\}  \nonumber  \\
&&+\frac{P}{8A\kappa^{3/2}}\Bigl\{ -8B\kappa X +8A\kappa Y +A\dot{\kappa}-2\dot{A}\kappa \Bigr\}\sin \{4\sqrt{\kappa}(Q-\frac{\partial s_{1}(P,t)}{\partial P})\}  \nonumber  \\
&&-\frac{P}{2}~\Bigl(Q-\frac{\partial s_{1}(P,t)}{\partial P} \Bigr)~\frac{d}{d t}\log \kappa.
\end{eqnarray}
 Hamilton's equations for $Q$ and $P$ are given by differentiating $K_{1}(Q,P,t)$:\begin{equation}
\dot{Q}=\frac{\partial K_{1}}{\partial P},~~~\dot{P}=-\frac{\partial K_{1}}{\partial Q}.\end{equation}
Thus we have obtained all necessary  formulae for the canonical transformation $(p,q) \to (P,Q)$, transforming  $(p,q)$ to the new canonical variables $(P,Q)$, where $P$ is defined by a  bilinear form in $p$ and $q$   with  arbitrary coefficient functions $A(t), B(t)$ and $C(t)$ of $t$ .

Next we consider a  special case of the canonical transformation such that the transformed Hamiltonian $K_{1}$ does not involve the canonical coordinate  Q, that is, the  transformation in which $Q$ becomes a cyclic coordinate. Now, for  $Q$ to be a cyclic coordinate, it is necessary and sufficient, as seen from (12),  that $A(t),B(t)$ and $C(t)$ satisfy the following relations:
\begin{eqnarray}
&&B\dot{A}-A\dot{B}+2B^{2}X -4ABY +2A^{2}Z -2\kappa X =0,  \\
&&-A\dot{\kappa}+2\dot{A}\kappa+8B\kappa X -8A\kappa Y =0,  \\
&&\dot{\kappa}=\dot{A}C+A\dot{C}-2B\dot{B}=0
\end{eqnarray}

From  (16) we see that $\kappa$ is a constant, and from (15) and (16)
 \begin{equation}
\dot{A} =-4\bigl(B X -A Y \bigr). 
\end{equation}
By substituting (17) into (14) we have
\begin{equation}
\dot{B}=\frac{2}{A}\{A^{2}Z -B^{2}X -\kappa X \},
\end{equation}
and from (15), (17) and (18) we have a second order, nonlinear differential equation for $A(t)$\cite{HRL}:
\begin{equation}
\ddot{A}-\frac{1}{2}\frac{\dot{A}^{2}}{A}-\frac{\dot{X}}{X}\dot{A}+2\Omega A-8X^{2}\frac{\kappa}{A}=0. 
\end{equation}
It is then easy to show that $B(t)$ and $C(t)$ can be expressed in terms of $A(t)$ as follows:
\begin{equation}
 B=\frac{4A Y -\dot{A} }{4X },\hspace{1cm} C=\frac{\kappa+B^{2}}{A}=\frac{\kappa}{A}+\frac{1}{16AX^{2}}(4AY-\dot{A})^{2} .
\end{equation}

That is to say, once  we know $A(t)$ as a solution of (19), $B(t)$ and $C(t)$ are given by (20) in terms of $A(t)$ . In this case the Hamiltonian $K_{1}(t)$ becomes  
\begin{equation}
K_{1}(P,t)=\frac{X }{A }~P+ \frac{\partial s_{1}(P,t)}{\partial t},
\end{equation}
which does not involve the coordinate variable $Q$ ($Q$ is a cyclic coordinate),  so that Hamilton's equation for $P$ becomes  
\begin{equation}
\dot{P}=-\frac{\partial K_{1}(P,t)}{\partial Q}=0.
\end{equation}
From (22) we see that the transformed canonical momentum $P$ given by (4), where the coefficients $A(t),B(t)$ and $C(t)$  satisfy (19) and (20), is a constant of motion.  It can then be shown that $P$ written in terms of canonical variables $(q,p)$ satisfies the following equation of motion
\begin{equation}
\dot{P}=\{P,H(t)\}+\frac{\partial P}{\partial t}=0.
\end{equation}

The transformed Hamiltonian $K_{1}$ given by (21) still remains undetermined because of  the presence of $\partial s_{1}(P,t)/\partial t$. Let us now consider the following  two cases ;~ i) $s_{1}(P,t)=0$, and  ii) $s_{1}(P,t)=-P \int_{0}^{t} \frac{X(t^{\prime})}{A(t^{\prime})}d t^{\prime}$.

i) When $s_{1}(P,t)=0$,  $K_{1}(P,t)$ becomes  
\begin{equation}
K_{1}(P,t)=\frac{X(t)}{A(t)}P.
\end{equation}
The canonical coordinate $Q$ then satisfies the equation of motion
\begin{equation}
\dot{Q}=\frac{X(t)}{A(t)},
\end{equation}
so that  
\begin{equation}
Q(t)=\int^{t} \frac{X(t^{\prime})}{A(t^{\prime})} d t^{\prime}.  
\end{equation}
By substituting (26) into (9) and (10) it is found that $q$ and p are given as
\begin{eqnarray}
&& q=\sqrt{\frac{AP}{\kappa}}\sin \bigl(\varphi(t)+\varphi_{0}\bigr),  \\
&& p=\sqrt{\frac{P}{A}}\Bigl\{-\frac{B}{\sqrt{\kappa}}\sin \bigl(\varphi(t)+\varphi_{0}\bigr)+\cos \bigl(\varphi(t)+\varphi_{0}\bigr) \Bigr\},  \nonumber \\
\end{eqnarray}
where $\varphi_{0}$ is a constant and $\varphi(t)$ is  defined by
\begin{equation}
\varphi(t) \equiv 2\sqrt{\kappa}\int_{0} ^{t} \frac{X(t^{\prime}) }{A(t^{\prime})}~d t^{\prime}.
\end{equation}
That is, (27) gives a formal solution  to the equation of motion (2) for $q$,  and (28) gives the corresponding expression for  $p$ .

In the special case where the coefficients $X,Y$ and $Z$ are all constants such that $XZ-Y^{2} > 0$,  it can be shown that
\begin{equation}
A=\sqrt{\frac{\kappa}{XZ-Y^{2}}}X,~~~B=\sqrt{\frac{\kappa}{XZ-Y^{2}}}Y, ~~{\rm and} ~~C=\sqrt{\frac{\kappa}{XZ-Y^{2}}}Z
\end{equation}
satisfy (19) and (20), and that $P$ and $Q$ become 
\begin{eqnarray}
&& P=\sqrt{\frac{\kappa}{XZ-Y^{2}}}\Bigl(X p^{2}+2Y pq+ Z q^{2}\Bigr)=\frac{1}{\nu}H,  \\
&& Q=\nu t+{\rm const.},
\end{eqnarray}
where $\nu$, the frequency of the system, is defined by $\nu=\sqrt{(XZ-Y^{2})/\kappa}$. Thus we see that $P$ and $Q$ given, respectively, by (31) and  (32) are nothing but the action and angle variables of this system, and that the transformation (9) and (10) with $s_{1}(P,t)=0$ is the canonical transformation leading to such  variables.

On the other hand, $P$ and $Q$ for the general cases, in which  $H(t)$ depends on $t$ explicitly, may be regarded as a generalization of  action and angle variables, respectively; however, the latter $Q(t)$ given by (26) is not, in general, a linear function in $t$. Thus,  the transformation considered above is a generalization of the canonical transformation to the action and angle variables.

ii) When $s_{1}$ is chosen to be
\begin{equation}
s_{1}(P,t)=-P \int_{0}^{t} \frac{X(t^{\prime})}{A(t^{\prime})} d t^{\prime},
\end{equation}
$K_{1}$ becomes  zero, and then $Q$ and $P$ are both constants of motion. Here the generating function of this transformation takes the form
\begin{equation}
S_{1}=-\frac{B }{2A }q^{2}+\frac{\sqrt{\kappa}}{2A }\Bigl[q\sqrt{\frac{A P}{\kappa}-q^{2}}+\frac{A P}{\kappa}\sin^{-1}(\sqrt{\frac{\kappa}{A P}}~q)\Bigr]-P\int_{0}^{t}\frac{X(t^{\prime})}{A(t^{\prime})}d t^{\prime},
\end{equation}
and  satisfies the Hamilton-Jacobi equation. By substituting  (33) into
(9) and (10), we can show that the formal solution for $q$ and $p$ are given in the same form as (27) and (28), respectively.

\subsection{Canonical transformation 2} 
 
In this subsection we consider another type of   canonical transformations  $(p,q) \to (\gamma, \eta)$, where  $\gamma$ is defined as a linear combination of $p$ and $q$:
\begin{equation}
\gamma \equiv f(t) p+ g(t) q.
\end{equation}  
with  $f(t)$ and $g(t)$ being some suitable real  functions of $t$. The differential equation for the generating function $S_{2}(q,\gamma,t)$ of the canonical transformation is given by 
\begin{equation}
p=\frac{\partial S_{2}(q,\gamma,t)}{\partial q}=\frac{\gamma}{f(t)}-\frac{g(t)}{f(t)}q, 
\end{equation} 
which after integration leads to
\begin{equation}
S_{2}(q,\gamma,t)=\frac{\gamma}{f(t)}q-\frac{g(t)}{2 f(t)}~q^{2}+s_{2}(\gamma,t),\end{equation} 
where $s_{2}(\gamma,t)$ is an arbitrary function of $\gamma$ and $t$. 

The canonical coordinate $\eta$,  conjugate to $\gamma$, is given by
\begin{equation}
\eta = \frac{\partial S_{2}(q,\gamma,t)}{\partial \gamma}=\frac{q}{f(t)}+\frac{\partial s_{2}(\gamma,t)}{\partial \gamma}.
\end{equation}
Then $q$ and $p$ are expressed in terms of $\eta$ and $\gamma$ as 
\begin{equation}
q=f(t)\Bigl(\eta-\frac{\partial s_{2}(\gamma,t)}{\partial \gamma}\Bigr),~~~~p=\frac{\gamma}{f(t)}-g(t)\Bigl(\eta-\frac{\partial s_{2}(\gamma,t)}{\partial \gamma}\Bigr).
\end{equation}
Eqs. (35),(38) and (39) prescribe the   canonical transformations 2.

By this canonical transformation the   Hamiltonian is transformed to $K_{2}(\eta,\gamma,t)$ : 
\begin{eqnarray}
K_{2}(\eta, \gamma, t)&=& \frac{X }{f^{2} }\gamma^{2}+\Bigl(-\frac{\dot{f} }{f }-2X \frac{g }{f }+2Y\Bigr)\gamma \Bigl(\eta-\frac{\partial s_{2}(\gamma,t)}{\partial \gamma}\Bigr) \nonumber  \\
&+&\Bigl(X g^{2} -2Y f g +Z f^{2} -\frac{\dot{g} f -g \dot{f} }{2}\Bigr)\Bigl(\eta-\frac{\partial s_{2}(\gamma,t)}{\partial \gamma}\Bigr)^{2}+\frac{\partial s_{2}(\gamma,t)}{\partial t}. \nonumber  \\
&&
\end{eqnarray}
This Hamiltonian depends on $\eta$ as well as $\gamma $ and $t$, hence $\eta$ in general is not a cyclic coordinate.
 
In the following we consider the case in which the   variable $\eta$ becomes a  cyclic coordinate. From (40) we see that the necessary and sufficient conditions for  $\eta$ to be such a coordinate are given by the differential equations:
\begin{equation}
\dot{f} =2Y f -2X g,~~~~~\dot{g} =2Z f -2Y g .
\end{equation}  
From (41) we have
\begin{equation}
\ddot{f} -\frac{\dot{X} }{X }\dot{f} +\Omega f =0,
\end{equation}
which is of the same form as the equation of motion (2), and
\begin{equation}
g =\frac{2f Y -\dot{f} }{2X }.
\end{equation}
When $f(t)$ and $g(t)$ satisfy, respectively,  (42) and (43), the transformed Hamiltonian $K_{2}$ becomes  
\begin{equation}
K_{2}=\frac{X }{f^{2} }\gamma^{2}+\frac{\partial s_{2}(\gamma,t)}{\partial t},
\end{equation}
which does not depend on $\eta$.  Hence $\eta$ is a cyclic coordinate,  and the canonical momentum $ \gamma$ becomes   a constant ($\dot{\gamma}=0$). 

As for the undetermined function  $s_{2}(\gamma,t)$ in $K_{2}$, we now consider  two cases :  i) $s_{2}(\gamma,t)=0$, and  ii) $s_{2}(\gamma,t)=-\gamma^{2}\int_{0}^{t} \frac{X(t^{\prime})}{f^{2}(t^{\prime})}d t^{\prime}$. 

i)When $s_{2}(\gamma,t)=0$, $K_{2}=\gamma^{2}~\bigl(X(t)/f^{2}(t)\bigr)$, and $\dot{\eta}=\partial K_{2}/\partial \gamma =2\gamma~\bigl(X(t)/f^{2}(t)\bigr) $.  Then $\eta$ is given by
\begin{equation}
\eta =2\gamma \tilde{\varphi}(t)+\eta_{0}, ~~~~\tilde{\varphi}(t) \equiv  \int_{0}^{t} \frac{X(t^{\prime})}{f^{2}(t^{\prime})}d t^{\prime} ,
\end{equation}
and $q$ is expressed in terms of $f(t)$ as
\begin{equation}
q=f \Bigl(2\gamma \tilde{\varphi}(t) +\eta_{0}\Bigr).
\end{equation}
This provides  another expression for the exact solution of (3),   obtained via the canonical transformation generated by $S_{2}(\gamma,q,t)$   with the vanishing $s_{2}(\gamma,t)$. 

ii) When $s_{2}(\gamma,t)=-\gamma^{2} \tilde{\varphi}(t)$, $K_{2}$ vanishes ($K_{2}=0$), and then both the canonical variables $\gamma$ and $\eta$ become   constants . In this case, the generating function $S_{2}$ is given by
\begin{equation}
S_{2}(q,\gamma,t)=\frac{\gamma}{f }q-\frac{g }{2f }q^{2}-\gamma^{2} \tilde{\varphi}(t),
\end{equation}
which satisfies the Hamilton-Jacobi equation, and   (38) and (35) lead to
\begin{equation}
\eta= -2f  \tilde{\varphi}(t) p+\Bigl(\frac{1}{f }-2g  \tilde{\varphi}(t)\Bigr)q.
\end{equation}

\subsection{The relation between $A(t)$ and $f(t)$}

In the above we have considered two kinds of canonical transformations generated by   $S_{1}(q,P,t)$ and $S_{2}(q,\gamma,t)$:  the one generated by $S_{1}(q,P,t)$   transforms $(p,q)$ to $(P,Q)$, whereas the one  generated by $S_{2}(q,\gamma,t)$  transforms $(p,q)$ to $(\gamma, \eta)$. The transformed momentum $P$ is a constant of motion, being   written in a bilinear form of $p$ and $q$. On the other hand,  $\gamma$ which is another constant of motion is given as a linear combination of $p$ and $q$. It has been shown that $P$ is expressed in terms of $A(t)$ ( as well as  $B(t)$ and $C(t)$ ) which satisfies the nonlinear differential equation (19), whereas   $\gamma$ is written in terms of $f(t)$ ( as well as   $g(t)$) which satisfies the linear differential equation (42).
We can now prove the following mathematical theorem. 

For any of two linearly independent solutions $f_{1}(t)$ and $f_{2}(t)$ of (42), there holds the relation
\begin{equation}
 \bigl(\dot{f}_{1}(t) f_{2}(t)-f_{1}(t) \dot{f}_{2}(t)\bigr)={\rm const.}\times X(t),
\end{equation}
where without loss of generality the above const. may be taken to be unity. Then, the expression 
\begin{equation}
A(t)=k_{1}\bigl(f_{1}^{2}(t)+f_{2}^{2}(t) \bigr)+k_{2}\bigl(f_{1}^{2}(t)-f_{2}^{2}(t)\bigr) +2k_{3} f_{1}(t)f_{2}(t),
\end{equation}
provides a solution of (19), with  $k_{1},k_{2}$ and $k_{3}$ being  constants satisfying the relation $(k_{1}^{2}-k_{2}^{2}-k_{3}^{2})=4\kappa$.

 The proof of (49) is straightforward, and that of (50) is not difficult when (49) is taken into account.
 Eq. (50) is very interesting in that a  solution of the nonlinear differential equation (19) can be expressed in terms of two linearly independent solutions of the linear differential equation (42). The reason for this is basically that $P$ of (4) is a kind of product of two $\gamma$'s of (35).

Let $f(t)$ be a solution of (42) which satisfies the initial condition $f(0)\ne 0$. We can  then   show that $f(t)\tilde{\varphi}(t)$ satisfies (42) with $f(0)\tilde{\varphi}(0)=0$. This means that if we take $f_{1}(t)=f(t)$ and $f_{2}(t)=-f(t)\tilde{\varphi}(t)$, $f_{1}(t)$ and $f_{2}(t)$ are two linearly independent solutions of (42) which satisfy the relation  $\bigl(\dot{f}_{1}(t) f_{2}(t)-f_{1}(t) \dot{f}_{2}(t)\bigr)= X(t)$. Substituting these $f_{1}(t)$ and $f_{2}(t)$ into (50) we have
\begin{equation}
A(t)=\{(k_{1}+k_{2})-2k_{3}\tilde{\varphi}(t)+(k_{1}-k_{2})\tilde{\varphi}^{2}(t)\} f^{2}(t).
\end{equation}

With $A(t)$ given by (51) we can now express $\varphi(t)$ in terms of $\tilde{\varphi}(t)$ as
\begin{eqnarray}
\varphi(t)&=& 2\sqrt{\kappa}\int _{0}^{t} \frac{X(t^{\prime})}{A(t^{\prime})}d t^{\prime}=2\sqrt{\kappa}\int _{0}^{t}\frac{X(t^{\prime})}{f^{2}(t^{\prime})}\times \frac{d t^{\prime}}{k_{1}+k_{2}-2k_{3}\tilde{\varphi}(t^{\prime})+(k_{1}-k_{2})\tilde{\varphi}^{2}(t^{\prime})}  \nonumber  \\
&=&2\sqrt{\kappa}\int _{0}^{\tilde{\varphi}(t)}\frac{d\tilde{\varphi}(t^{\prime})}{k_{1}+k_{2}-2k_{3}\tilde{\varphi}(t^{\prime})+(k_{1}-k_{2})\tilde{\varphi}^{2}(t^{\prime})}  \nonumber  \\
&=&\tan^{-1}\Bigl\{\frac{(k_{1}-k_{2})\tilde{\varphi}(t)-k_{3}}{2\sqrt{\kappa}}\Bigr\}+\tan^{-1}\Bigl\{\frac{k_{3}}{2\sqrt{\kappa}}\Bigr\},
\end{eqnarray}
so that  
\begin{equation}
\cos \varphi(t)=\frac{k_{1}+k_{2}-k_{3}\tilde{\varphi}(t)}{\sqrt{k_{1}+k_{2}-2k_{3}\tilde{\varphi}(t)+(k_{1}-k_{2})\tilde{\varphi}^{2}(t)}\times\sqrt{k_{1}+k_{2}}},
\end{equation}
and
\begin{equation}
\sin \varphi(t)=\frac{2\sqrt{\kappa}\tilde{\varphi}(t)}{\sqrt{k_{1}+k_{2}-2k_{3}\tilde{\varphi}(t)+(k_{1}-k_{2})\tilde{\varphi}^{2}(t)}\times\sqrt{k_{1}+k_{2}}}.
\end{equation}

From (51), (53) and (54), the expression (27) can be rewritten in the form
\begin{equation}
q=\sqrt{\frac{P}{(k_{1}+k_{2}) \kappa}}~\{(2\sqrt{\kappa}\cos \varphi_{0}-k_{3}\sin \varphi_{0})\tilde{\varphi}(t)+(k_{1}+k_{2})\sin \varphi_{0}\}  f(t),
\end{equation}
which is equivalent to (46), provided that the parameters $\gamma$ and $\eta_{0}$ are assumed to be
\begin{equation}
2\gamma \equiv \sqrt{\frac{P}{(k_{1}+k_{2})\kappa}} (2\sqrt{\kappa}\cos \varphi_{0}-k_{3}\sin \varphi_{0}),~~\eta_{0}\equiv \sqrt{\frac{P}{(k_{1}+k_{2})\kappa}}(k_{1}+k_{2})\sin \varphi_{0}.
\end{equation}
Thus we have  shown that two kinds of canonical transformations generated by $S_{1}(q,P,t)$ and $S_{2}(q,\gamma,t)$ lead to the same   classical solution to the equation of motion (2).

\section{Quantum Solutions } 
 
In this section  we discuss, in the Schr$\ddot{{\rm o}}$dinger picture, the quantum theory of the system with $H(t)$ given by 
\begin{equation}
H(t)=X(t) p^{2}+Y(t)(pq+qp)+Z(t) q^{2},
\end{equation}
which is the quantum version of (1).

\subsection{ Formal solutions to the Schr$\ddot{{\rm o}}$dinger equation}

In a previous paper \cite{OK}, we have shown that formal solutions $|\phi(t)\bigr>$ of the Schr$\ddot{{\rm o}}$dinger equation
\begin{equation}
i\hbar \frac{\partial}{\partial t}|\phi(t)\bigr>=H(t)|\phi(t)\bigr>,
\end{equation}
can be written in the form
\begin{equation}
|\phi(t)\bigr>=\sum_{n} c_{n}(t) \exp \Bigl[\frac{i}{\hbar}\int_{0}^{t} 
 \theta_{n}(t^{\prime}) d t^{\prime} \Bigr]|n;t \bigr>,
\end{equation}
where $|n;t\bigr>$ represents the eigenstate of a hermitian operator $\Lambda(t)$ with  eigenvalue $\lambda_{n}$. Here we assume for simplicity  that the eigenstates are non-degenerate, and the set of $|n;t\bigr>$ forms an ortho-normal complete set (for the degenerate cases, see ref.4): 
\begin{equation}
\begin{array}{l}
\Lambda(t) |n;t\bigr>=\lambda_{n}(t)|n;t\bigr>, ~~\bigl<n;t|m;t\bigr>=\delta_{nm},  \\
\sum_{n}|n;t\bigr>\bigl<n;t|={\rm I}.
\end{array}
\end{equation}
The phase function $\theta_{n}(t)$ is defined by
\begin{equation}
\theta_{n}(t)\equiv \bigl<n;t|\bigl(i\hbar\frac{\partial}{\partial t}-H(t)\bigr)|n;t\bigr>=\theta^{\ast}(t).
\end{equation}
In this connection let us note that for the case when $\Lambda(\tau)=\Lambda(0)$, or $|n;\tau\Bigr> \propto|n;0\bigr>$ the time integrals $(-1/\hbar)\int_{0}^{\tau} dt \bigl<n;t|H(t)|n;t\bigr>$ and $\int_{0}^{\tau} dt \bigl<n;t|i\partial/\partial t|n;t\bigr>$ provide, respectively, the so-called dynamical and geometrical phases of the states $|n;t\bigr>$ \cite{MVB,YA-JA} .

The coefficient $c_{n}(t)$ in (59) satisfies the differential equation
\begin{equation}
\frac{d c_{n}(t)}{d t}=\sum_{n^{\prime} \ne n}~\exp\Bigl[\frac{i}{\hbar}\int_{0}^{t} \Bigl((\theta_{n^{\prime}}(t^{\prime})-\theta_{n}(t^{\prime})\Bigr)d t^{\prime}\Bigr]\times \frac{\Bigl<n;t|\frac{{\cal D} \Lambda(t)}{{\cal D} t}|n^{\prime};t\Bigr>}{\bigl(\lambda_{n}(t)-\lambda_{n^{\prime}}(t)\bigr)} ~c_{n^{\prime}}(t),
\end{equation}
where ${\cal D}\Lambda (t)/{\cal D}t$ denotes
\begin{equation}
\frac{{\cal D}\Lambda(t)}{{\cal D} t}\equiv \frac{\partial \Lambda(t)}{\partial t}+\frac{1}{i \hbar}[\Lambda(t),H(t)].
\end{equation}
From (62) and (63),  we see that when $\Lambda(t)$ satisfies the relation 
\begin{equation}
  \frac{{\cal D} \Lambda(t)}{{\cal D} t} =0,
\end{equation}
$\Lambda(t)$ corresponds to a conserved quantity, and all eigenvalues $\lambda_{n}(t)$ do not depend on $t$: $\lambda_{n}(t)=\lambda_{n} (={\rm const})$. Incidentally, (64) is the quantum version of the classical equation (23), but should not be regarded here as  the equation of motion for the operator $\Lambda (t)$, since all operators concerned are those in the Schr$\ddot{{\rm o}}$dinger picture.

 Now for such cases all the coefficients $c_{n}(t)$ become constants, and the formal solution  $|\phi(t)\bigr>$ is reduced to \cite{OK,HLJ-WBR}
\begin{equation}
|\phi(t)\bigr>=\sum_{n} c_{n} \exp \Bigl[\frac{i}{\hbar}\int_{0}^{t} 
 \theta_{n}(t^{\prime}) d t^{\prime} \Bigr]|n;t \bigr>.
\end{equation} 
In this case transitions between eigenstates with different eigenvalues do not occur, that is, in the course of time the state $|n;t\bigr>$ keeps its identity specified by n.

\subsection{Formal solutions for a generalized oscillator}
We now  consider a hermitian operator $\Lambda(t)$ such as
\begin{equation}
\Lambda(t)=A(t) p^{2}+B(t)(p q+q p)+C(t)q^{2},
\end{equation}
which corresponds to the classical expression (4). It is then easy to see that $\Lambda(t)$ satisfies the relation (64), provided that $A(t), B(t)$ and $C(t)$ are solutions of (19) and (20) . Hereafter we shall restrict ourselves again to the case $\kappa>0$.

In order to obtain eigenstates of $\Lambda(t)$, we introduce new operators $a(t)$ and $a^{\dagger}(t)$ such as 
\begin{eqnarray}
&& a(t)=\frac{1}{\sqrt{2\hbar\sqrt{\kappa}A(t)}}\Bigl\{(\sqrt{\kappa}+iB(t))q+i A(t) p \Bigr\},  \\
&& a^{\dagger}(t)=\frac{1}{\sqrt{2\hbar\sqrt{\kappa}A(t)}}\Bigl\{(\sqrt{\kappa}-iB(t))q-i A(t) p \Bigr\},
\end{eqnarray}
which are operators depending on time $t$ explicitly,  and  which then satisfy the commutation relation
\begin{equation}
[a(t), a^{\dagger}(t)]=1.
\end{equation}
In terms of $a(t)$ and $a^{\dagger}(t)$, $p$ and $q$ are written as
\begin{eqnarray}
&&p=\sqrt{\frac{\hbar}{2A(t)\sqrt{\kappa}}}~\bigl\{-(B(t)+i\sqrt{\kappa})a(t) +(-B(t)+i\sqrt{\kappa})a^{\dagger}(t)\bigr\},  \\
&&q=\sqrt{\frac{\hbar A(t)}{2\sqrt{\kappa}}}~\bigl\{a(t) + a^{\dagger}(t)\bigr\}.
\end{eqnarray}
Substituting (70) and (71) into (66), $\Lambda(t)$ is rewritten  as 
\begin{equation}
\Lambda(t)=2\hbar \sqrt{\kappa}\Bigl(N(t)+\frac{1}{2}\Bigr),
\end{equation}
where $N(t)$ is the time-dependent hermitian operator defined by $N(t)\equiv a^{\dagger}(t) a(t)$.

The operator $\Lambda(t)$ and $N(t)$ are simultaneously diagonalized with respect to the set of  eigenstates $|n;t\bigr>$ : 
\begin{equation}
|n;t\bigr> \equiv \frac{1}{\sqrt{n!}}\bigl(a^{\dagger}(t)\bigr)^{n}~|0;t\bigr>,~~~(n=0,1,2,\cdots),
\end{equation}
where $|0;t \bigr>$ is defined by
\begin{equation}
a(t)~|0;t\bigr>=0.
\end{equation} 
These eigenstates $|n;t\bigr>$ satisfy the following relations:
\begin{equation}
\begin{array}{ll}
N(t)|n;t\bigr>=n|n;t\bigr>,&   \Lambda(t)|n;t\bigr>= 2\hbar \sqrt{\kappa}\Bigl(n+\frac{1}{2}\Bigr),  \\
\sum_{n} |n;t \bigr>\bigl<n;t|={\rm I}, & \bigl<n;t|m;t\bigr>=\delta_{nm}.
\end{array}
\end{equation} 

Let $|q\bigr>$'s be eigenstates of $\hat{q}$: $\hat{q}|q\bigr>=q|q\bigr>$. (Hereafter the symbol $~\hat{}~$ is used for operators whenever neccesary.) Then, the $q$- or $configuration-space ~ representation$ $\bigl<q|n;t\bigr>$ of  $|n;t\bigr>$,   can be obtained from (67),(68) and (73) as follows . To this end, let us first consider   $\bigl<q|0;t\bigr>$ . From (67) and (74) we have
\begin{equation}
\bigl<q|a(t)|0;t\bigr>=\frac{1}{\sqrt{2\hbar A(t)\sqrt{\kappa}}}\Bigl[\bigl(\sqrt{\kappa}+iB(t)\bigr)q +\hbar A(t) \frac{d}{d q} \Bigr]\bigl<q|0;t\bigr>=0,
\end{equation}
so that  $\bigl<q|0;t\bigr>$ is found to be
\begin{equation}
\bigl<q|0;t\bigr>=\Bigl(\frac{\sqrt{\kappa}}{\hbar \pi A(t)}\Bigr)^{1/4}\exp\bigl\{-\frac{(\sqrt{\kappa}+iB(t))}{2\hbar A(t)}q^{2}\bigr\}.
\end{equation}
It should be noticed that $\bigl<q|0;t\bigr>$ depends on $t$:  this $t$-dependence, of course,   has nothing to do with the one arising from the Schr$\ddot{{\rm o}}$dinger equation.  

Next from (68),  $a^{\dagger}(t)$ can be written as
\begin{eqnarray}
a^{\dagger}(t)&=&\frac{1}{\sqrt{2\hbar A(t)\sqrt{\kappa}}}\Bigl\{(\sqrt{\kappa}-iB(t))q-\hbar A(t) \frac{d}{d q} \Bigr\}  \nonumber  \\
&=&-\sqrt{\frac{\hbar A(t)}{2\sqrt{\kappa}}}~\exp\Bigl\{\frac{(\sqrt{\kappa}-iB(t))}{2\hbar A(t)}q^{2}\Bigr\}\frac{d}{d q}~\exp\Bigl\{-\frac{(\sqrt{\kappa}-iB(t))}{2\hbar A(t)}q^{2}\Bigr\},
\end{eqnarray}
so that $\bigl(a^{\dagger}(t)\bigr)^{n}$ becomes 
\begin{equation}
\bigl(a^{\dagger}(t)\bigr)^{n}=\Bigl(-\sqrt{\frac{\hbar A(t)}{2\sqrt{\kappa}}}~\Bigr)^{n}~\exp\Bigl\{\frac{(\sqrt{\kappa}-iB(t))}{2\hbar A(t)}q^{2}\Bigr\}~\frac{d^{n}}{d q^{n}}~\exp\Bigl\{-\frac{(\sqrt{\kappa}-iB(t))}{2\hbar A(t)}q^{2}\Bigr\}.
\end{equation} 
Thus we have
\begin{eqnarray}
\bigl<q|n;t\bigr>&=&\frac{1}{\sqrt{n!}}\bigl<q|(a^{\dagger}(t))^{n}|0;t\bigr>  \nonumber  \\
&=&\sqrt{\frac{\xi(t)}{n!\sqrt{2\pi}}}~\exp\{-\frac{\xi^{2}(t)}{4}q^{2}-\frac{iB(t)}{2\hbar A(t)}q^{2}\}~H_{n}(\xi q),
\end{eqnarray}
where $\xi(t) \equiv \sqrt{2\sqrt{\kappa}/\bigl(\hbar A(t)\bigr)}$, and the Hermite polynomial $H_{n}(x)$ is given by
\begin{equation}
H_{n}(x) \equiv (-1)^{n} e^{ x^{2}/2}\frac{d^{n}}{d x^{n}}e^{- x^{2}/2}.
\end{equation} 

We next proceed to calculate the phase function $\theta_{n}(t)$. The Hamiltonian $H(t)$ given by (57) is expressed in terms of $a(t)$ and $a^{\dagger}(t)$ as 
\begin{eqnarray}
H&=&\frac{\hbar}{2\sqrt{\kappa}}\Bigl[\bigl\{\frac{X(t)}{A(t)}(B^{2}(t)+2i\sqrt{\kappa}B(t)-\kappa)-2Y(t)(B(t)+i\sqrt{\kappa})+A(t)Z(t)\bigr\}a^{2}(t)  \nonumber  \\
&&~~~~~+\bigl\{\frac{X(t)}{A(t)}(B^{2}(t)-2i\sqrt{\kappa}B(t)-\kappa)+2Y(t)(-B(t)+i\sqrt{\kappa})+A(t)Z(t)\bigr\}(a^{\dagger}(t))^{2}  \nonumber  \\
&&~~~~~~+\bigl\{\frac{X(t)}{A(t)}(B^{2}(t)+\kappa)-2Y(t)B(t)+A(t)Z(t)\bigr\}\bigl(a(t)a^{\dagger}(t)+a^{\dagger}(t)a(t)\bigr)\Bigr].
\end{eqnarray}
A straightforward calculation then shows that  
\begin{equation}
\bigl<n;t|H(t)|n;t \bigr>=(n+\frac{1}{2})\frac{\hbar}{\sqrt{\kappa}}\Bigl\{\frac{X(t)}{A(t)}(B^{2}(t)+\kappa)-2Y(t)B(t)+A(t)Z(t)\Bigr\},
\end{equation}
and by use of (80) 
\begin{equation}
\bigl<n;t|i\hbar\frac{\partial}{\partial t}|n;t\bigr>=\int d q \bigl<n;t|q\bigr>\Bigl(i\hbar\frac{\partial}{\partial t}\Bigr)\bigl<q|n;t\bigr>=(n+\frac{1}{2})\frac{\hbar}{2\sqrt{\kappa}}\frac{\dot{B}(t)A(t)-B(t)\dot{A}(t)}{A(t)},
\end{equation}
hence $\theta_{n}(t)$ turns out to be 
\begin{equation}
\theta_{n}(t)=\bigl<n;t|(i\hbar\frac{\partial}{\partial t}-H(t))|n;t\bigr>=-\Bigl(n+\frac{1}{2}\Bigr)\Bigl(\hbar\sqrt{\kappa}\Bigr)~\frac{2X(t)}{A(t)}.
\end{equation}

Thus  from (65), (80) and (85) the formal solution to the Schr$\ddot{{\rm o}}$dinger equation (58) is  given by
\begin{eqnarray}
\bigl<q|\phi(t)\bigr>&=& \sum_{n} c_{n} \sqrt{\frac{\xi(t)}{n!\sqrt{2\pi}}}~\exp\Bigl[-i(n+\frac{1}{2}) \varphi(t) \Bigr]~\exp\Bigl[-\frac{\xi^{2}(t)}{4}q^{2}-\frac{iB(t)}{2\hbar A(t)}q^{2}\Bigr]H_{n}\bigl(\xi(t) q\bigr),  \nonumber  \\
&&
\end{eqnarray}
where $\varphi(t)$ is the same function as that   defined by (29). 

Here, of great theoretical interest is the fact that the time-dependence of the classical and quantum solutions are both characterized by the same function $\varphi(t)$. As has been noticed,  this function arises from the (generalized) angle variable $Q(t)$ in the classical case,  and from the phase factor $\theta_{n}(t)$ in the quantum case. The fact that the time integral of $\theta_{n}(t)$ consists  of the dynamical and  geometrical phases indicates the importance of the role played by  the latter  phases in the quantum case.

\subsection{ Feynman kernels  }

As is clear from (65), the time evolution operator $U(t,0)$ of this system is given by
\begin{equation}
U(t,0)=\sum_{n}|n;t\bigl> \exp\Bigl[\frac{i}{\hbar}\int_{0}^{t} \theta_{n}(t^{\prime}) d t^{\prime} \Bigr]\bigl<n;0|.
\end{equation}
Then $\bigr<q_{1}|U(t,0)|q_{2}\bigr>$ is given as
\begin{eqnarray}
 \bigr<q_{1}|U(t,0)|q_{2}\bigr> &=& \sum_{n}\exp\bigl[\frac{i}{\hbar}\int_{0}^{t} \theta_{n}(t^{\prime}) d t^{\prime}\bigr]~\bigl<q_{1}|n;t\bigr>\bigl<n;0|q_{2}\bigr>  \nonumber  \\
&&  =\Bigl(\frac{\sqrt{\kappa}}{2\pi i\hbar \sin \varphi(t)}\Bigr)^{1/2}\Bigl(A(t)A(0)\Bigr)^{-1/4}\times \exp\Bigl[-\frac{i}{2\hbar}\Bigl(\frac{B(t)}{A(t)}q_{1}^{2}-\frac{B(0)}{A(0)}q_{2}^{2}\Bigr)\Bigr]  \nonumber  \\
&& \hspace{1.5cm}\times \exp  \Bigl[i \frac{(\xi^{2}(t) q_{1}^{2}+\xi^{2}(0) q_{2}^{2})\cos \varphi(t)-2\xi(t)\xi(0) q_{1}q_{2}}{4\sin\varphi(t) }\Bigr],    
\end{eqnarray}
thereby providing the Feynman kernel of this system. In deriving this, we have used the formula:
\begin{eqnarray}
&& \sum_{n=0}^{\infty} \frac{e^{-\beta(n+\frac{1}{2})}}{n!\sqrt{\pi}}H_{n}(\xi_{1}x_{1})H_{n}(\xi_{2}x_{2})\exp\Bigl\{-\frac{(\xi_{1}^{2}x_{1}^{2}+\xi_{2}^{2}x_{2}^{2})}{4}\Bigr\}  \nonumber  \\
&& \hspace{1cm} =\frac{1}{\sqrt{2\pi \sinh \beta}}\exp\Bigl\{-\frac{(\xi_{1}^{2}x_{1}^{2}+\xi_{2}^{2}x_{2}^{2})\cosh \beta -2\xi_{1} \xi_{2} x_{1}x_{2}}{4\sinh \beta}\Bigr\}.
\end{eqnarray}
 
\section{ Coherent States and the Classical Limit}

In this section we consider  eigenstates of the operator $a(t)$ introduced in (67) and discuss their classical limit . 
\subsection{Coherent states}

By $|\alpha;t\bigr>$  we denote the eigenstate of $a(t)$ with eigenvalue $\alpha$, or the so-called coherent state:   
\begin{equation}
a(t)|\alpha;t\bigr>=\alpha|\alpha;t\bigr>,
\end{equation}
which is  given as a linear combination of $|n;t\bigr>$'s such as
\begin{equation}
|\alpha;t\bigr>=e^{-\frac{|\alpha|^{2}}{2}}\sum_{n=0}^{\infty} \frac{\alpha^{n}}{\sqrt{n!}}|n;t\bigr>.
\end{equation}
Here  $\alpha$ is an arbitrary complex number, so that $a(t)$ and $a(t^{\prime}) ~(t \ne t^{\prime})$ can share the same, $t$-independent eigenvalue $\alpha$.  Since $a(t)$ depends on $t$ explicitly, the state $|\alpha;t\bigr>$ has   time-dependence  which comes in the above from that of $|n;t\bigr>$'s. We remark here that this time dependence of $|\alpha;t\bigr>$ should be distinguished from that of time evolution due to the Schr$\ddot{{\rm o}}$dinger  equation. 
 
On the other hand, the state which has  evolved from the state $|\alpha;0\bigr>$ at $t=0$ according to the Schr$\ddot{{\rm o}}$dinger  equation will be denoted hereafter by $|\phi(t)\bigr>_{\alpha}$, hence
\begin{equation}
|\phi(t)\bigr>_{\alpha}=U(t,0)|\alpha;0\bigr>.
\end{equation}
Substituting  (87) into (92) we have
\begin{equation}
|\phi(t)\bigr>_{\alpha}=\exp\Bigl[-\frac{|\alpha|^{2}}{2}-\frac{i}{2}\varphi(t)\Bigr]~~\sum_{n=0}^{\infty} \frac{(\tilde{\alpha}(t))^{n}}{\sqrt{n!}}|n;t\bigr>,
\end{equation}
where $\tilde{\alpha}(t)$ is defined by
\begin{equation}
\tilde{\alpha}(t)\equiv \alpha \exp\bigl(-i\varphi(t)\bigr).
\end{equation}

Although,  as noted above, the definitions of $|\alpha;t\bigr>$ and of $|\phi(t)\bigr>_{\alpha}$ are different,  it is also true  that  $|\phi(t)\bigr>_{\alpha}$  is another eigenstate of $a(t)$. This is seen as follows.  From (87) we have
\begin{eqnarray}
U^{-1}(t,0)a(t)U(t,0)&=&\sum_{n,m}e^{-\frac{i}{\hbar}\int_{0}^{t} \theta_{n}(t^{\prime})d t^{\prime}}|n;0\bigr>\bigl<n;t|a(t)|m;t\bigr>\bigl<m;0|e^{\frac{i}{\hbar}\int_{0}^{t} \theta_{m}(t^{\prime})d t^{\prime}}  \nonumber  \\
&&=\sum_{n=1}\sqrt{n}\exp\Bigl[-\frac{i}{\hbar}\int_{0}^{t}\{\theta_{n-1}(t^{\prime})-\theta_{n}(t^{\prime})\}d t^{\prime}\Bigr]|n-1;0\bigr>\bigl<n;0|.   \nonumber  \\
&&
\end{eqnarray}
Since $\theta_{n-1}(t)-\theta_{n}(t)=2\hbar \sqrt{\kappa} X(t)/A(t)$ from (85),   (95) can be rewritten as
\begin{equation}
U^{-1}(t,0)a(t)U(t,0)=\exp[-i\varphi(t)] \sum_{n=1}\sqrt{n}|n-1;0\bigr>\bigl<n;0|   =a(0)\exp[-i\varphi(t)] .
\end{equation}
Using (96) we have
\begin{eqnarray}
a(t)|\phi(t)\bigr>_{\alpha}&=&a(t)U(t,0)|\alpha;0\bigr>=U(t,0)U^{-1}(t,0)a(t)U(t,0)|\alpha;0\bigr> \nonumber  \\
&&=\exp[-i\varphi(t)] ~U(t,0) a(0) |\alpha;0\bigr> =\exp[-i\varphi(t) ] ~\alpha~ U(t,0)|\alpha;0\bigr>  \nonumber  \\
&&=\tilde{\alpha}(t)|\phi(t)\bigr>_{\alpha}.
\end{eqnarray}
That is to say, $|\phi(t)\bigr>_{\alpha}$ is not only a solution to the Schr$\ddot{{\rm o}}$dinger  equation but also the eigenastate of $a(t)$  with eigenvalue $\tilde{\alpha}(t)$ . This important result, in fact,  comes from the facts that $|\alpha;t\bigr>$ is an eigenstate of $a(t)$,  and that $a(t)$ and $a^{\dagger}(t)$ are operators which diagonalize the operator $\Lambda(t)$   satisfying the relation (64).

From (80) the $q$- representation $\bigl<q|\phi(t)\bigr>_{\alpha}$ of $|\phi(t)\bigr>_{\alpha}$ is found to be  
\begin{eqnarray}
\bigl<q|\phi(t)\bigr>_{\alpha}&=&\sqrt{\frac{\xi(t)}{\sqrt{2\pi}}}\exp\Bigl[-\frac{|\alpha|^{2}}{2}-\frac{i}{2}\varphi(t)-\frac{\xi^{2}(t)}{4}q^{2}-\frac{i}{2\hbar}\frac{B(t)}{A(t)}q^{2}\Bigr]  ~\sum_{n=0}^{\infty}\frac{(\tilde{\alpha}(t))^{n}}{n!}H_{n}(\xi q)  \nonumber  \\
&&=\sqrt{\frac{\xi(t)}{\sqrt{2\pi}}}\exp\Bigl[-\frac{|\alpha|^{2}}{2}-\frac{i}{2}\varphi(t)-\frac{\xi^{2}(t)}{4}q^{2} -\frac{i}{2\hbar}\frac{B(t)}{A(t)}q^{2} +\tilde{\alpha}(t)\xi(t)q-\frac{1}{2}(\tilde{\alpha})^{2}\Bigr] .
\end{eqnarray}
This is  the coherent state satisfying the Schr$\ddot{{\rm o}}$dinger equation,   which is also obtainable from (86) by putting  $c_{n}=\alpha^{n}\exp(-\frac{|\alpha|^{2}}{2})/\sqrt{n!}$ . 

In order to study  properties of this coherent state, we calculate the expectation values of $q,p$,  $q^{2}$ and $p^{2}$  for this state:
\begin{eqnarray}
&&\bigl<q\bigr>=\frac{\tilde{\alpha}(t)+\tilde{\alpha}^{\ast}(t)}{\xi(t)},  \\
&&\bigl<p\bigr>=\Bigl(\frac{\hbar \xi(t)}{2i}\Bigr) (\tilde{\alpha}(t)-\tilde{\alpha}^{\ast}(t))-\Bigl(\frac{B(t)} {A(t)}\Bigr)\frac{\tilde{\alpha}(t)+\tilde{\alpha}^{\ast}(t)}{\xi(t)} ,\\
&&\bigl<q^{2}\bigr>=\frac{1+(\tilde{\alpha}(t)+\tilde{\alpha}^{\ast}(t))^{2}}{\xi^{2}(t)},  \\
&&\bigl<p^{2}\bigr>=\frac{\hbar^{2}\xi^{2}(t)}{4}-\Bigl(\frac{\hbar^{2}\xi^{2}(t)}{4}\Bigr)(\tilde{\alpha}(t)-\tilde{\alpha}^{\ast}(t))^{2}+i\hbar\Bigl(\frac{ B(t)}{A(t)}\Bigr)(\tilde{\alpha}^{2}(t)-\tilde{\alpha}^{\ast 2}(t)) \nonumber  \\&& \hspace{3cm} +\Bigl(\frac{B^{2}(t)}{A^{2}(t)}\Bigr)\frac{ 1+(\tilde{\alpha}(t)+\tilde{\alpha}^{\ast}(t))^{2} }{\xi^{2}(t)}.   
\end{eqnarray}
The root-mean-square deviations $\triangle q$ of $q$ and $\triangle p$ of  $p$ for this state are then given by
\begin{eqnarray}
&&(\triangle q)^{2} \equiv \bigl<q^{2}\bigr>-\bigl<q\bigr>^{2}=\frac{1}{\xi^{2}(t)},  \\
&&(\triangle p)^{2} \equiv \bigl<p^{2}\bigr>-\bigl<p\bigr>^{2}=\frac{\hbar^{2}\xi^{2}(t)}{4}+\frac{B^{2}(t)}{A^{2}(t)\xi^{2}(t)},
\end{eqnarray}
thereby implying the uncertainty relation
\begin{equation}
(\triangle q)^{2}(\triangle p)^{2}=\frac{\hbar^{2}}{4}+\frac{B^{2}(t)}{A^{2}(t)\xi^{4}(t)}=\frac{\hbar^{2}}{4}\frac{A(t)C(t)}{\kappa}.
\end{equation}
From (105) we see that the coherent state (98) is not, in general (so far as  $B(t)\ne 0$), the state which has the minimum uncertainty .

\subsection{The classical limit of coherent states}
 Eq. (98) can be rewritten as
\begin{eqnarray}
\bigl<q|\phi(t)\bigr>_{\alpha}&=&\sqrt{\frac{\xi}{\sqrt{2\pi}}}\exp\Bigl[\Bigl\{-\frac{|\alpha|^{2}}{2}-\frac{\xi^{2}}{4}q^{2}-\frac{1}{2}|\alpha|^{2}\cos\Bigl(2\varphi(t)+2 \epsilon\Bigr)+|\alpha|\xi q\cos\Bigl( \varphi(t)+ \epsilon\Bigr)\Bigr\}  \nonumber  \\
&&~~+i \Bigl\{-\frac{1}{2}\varphi(t)-\frac{1}{2\hbar}\frac{B(t)}{A(t)}q^{2}+\frac{1}{2}|\alpha|^{2}\sin\Bigl(2\varphi(t)+2 \epsilon\Bigr)-|\alpha|\xi q\sin\Bigl(\varphi(t)+ \epsilon\Bigr)\Bigr\} \Bigr]. \nonumber  \\
&&
\end{eqnarray}
where $\alpha \equiv |\alpha|e^{-i\epsilon}$. 
Then we have
\begin{eqnarray}
|\bigl<q|\phi(t)\bigr>_{\alpha}|^{2}&=& \frac{\xi(t)}{\sqrt{2\pi}}\exp \Bigl\{-|\alpha|^{2}-\frac{\xi^{2}(t)}{2}q^{2}-|\alpha|^{2}\cos\Bigl(2\varphi(t)+2\epsilon \Bigr)+2|\alpha|\xi(t) q\cos\Bigl( \varphi(t)+\epsilon \Bigr)\Bigr\}  \nonumber  \\
&&=\frac{\xi}{\sqrt{2\pi}}\exp \Bigl[-\Bigl(\frac{\xi(t)}{\sqrt{2}}\Bigr)^{2}\Bigl\{q-2\frac{|\alpha|}{\xi(t)}\cos\Bigl(
\varphi(t)+ \epsilon\Bigr)\Bigr\}^{2} \Bigr].
\end{eqnarray}

We assume that $\alpha$ is given by $|\alpha|=\sqrt{2\sqrt{\kappa}/\hbar A}$,  then  $|\alpha|/\xi=\sqrt{AP/4\kappa}$.  In this case the limiting procedure $\hbar \to 0$ yields
\begin{equation}
\lim_{\hbar \to 0} |\alpha| \to \infty, ~~\lim_{\hbar \to 0} \xi \to \infty, 
\end{equation}
while keeping $|\alpha|/\xi$ as $\sqrt{A(t)P/4\kappa}$. Taking $\hbar \to 0$ in (107), we see that  
\begin{equation}
\lim_{\hbar \to 0}|\bigl<q|\phi(t)\bigr>_{\alpha}|^{2} =\delta\Bigl\{q-\sqrt{\frac{A(t)P}{\kappa}}\cos\Bigl(\varphi(t)+\epsilon \Bigr)\Bigr\},
\end{equation}
where we have used the formula
\begin{equation}
\lim_{\lambda \to \infty} \sqrt{\frac{\lambda}{\pi}}e^{-\lambda x^{2}}=\delta(x).
\end{equation}
In other words, the motion of the particle in the state $\bigl<q|\phi(t)\bigr>_{\alpha}$  approaches the classical orbit 
(27) in the   limit $\hbar \to 0 $ $(|\alpha| \to \infty $) for $\epsilon =\varphi_{0} +3\pi/2$.

The above result is interesting in the following : as seen in (105), $\bigl<q|\phi(t)\bigr>_{\alpha}$ does not have, in general, the minimum uncertainty, and nevertheless has  the right classical limit.

\section{Quantum solutions in the Hamilton-Jacobi picture}
As we have shown in 2.2, the coordinate variable $\eta$ given by (48) is a constant of motion, and the transformed Hamiltonian expressed in terms of $\eta$ and $\gamma$ (the canonical momentum conjugate to  $\eta$) vanishes. In this section we shall examine the corresponding situation in quantum mechanics, and call the resulting picture thereof the Hamilton-Jacobi (HJ) picture, in contrast to the Schr$\ddot{{\rm o}}$dinger picture.  

\subsection{The transformation  to the HJ picture}
An eigenstate $|\eta;t\bigr>$ of the operator $ \hat{\eta}$  with eigenvalue $\eta$    satisfies $ \hat{\eta} |\eta;t\bigr>=\eta |\eta;t\bigr>$ . It is noticed here that  the eigenstate $|\eta;t\bigr>$ depends on $t$,  but the eigenvalue $\eta$ does not, because as is easily seen, the operator $\hat{\eta}$ satisfies (64). Since $\hat{\gamma}$ is of the form (48), the $q$-representation $\bigl<q|\eta;t\bigr>$ of $|\eta;t\bigr>$ can be obtained from the differential equation

\begin{equation}
\bigl<q| \hat{\eta}|\eta;t\bigr>=\Bigl[2i\hbar f(t)\tilde{\varphi}(t) \frac{\partial}{\partial q}+\Bigl(\frac{1}{f(t)}-2g(t)\tilde{\varphi}(t)\Bigr)q\Bigr]\bigl<q|\eta;t\bigr>=\eta\bigl<q|\eta;t\bigr>.
\end{equation}
Integrating (111) we   have 
\begin{equation}
\bigl<q|\eta;t\bigr>=\Theta(\eta)~\bigl(4\pi \hbar f(t) \tilde{\varphi}(t)\bigr)^{-1/2}\exp\Bigl[-\frac{i}{\hbar}\frac{\eta q +(g(t)\tilde{\varphi}(t)-\frac{1}{2f(t)})q^{2}}{2f(t)\tilde{\varphi}(t)}\Bigr]
\end{equation}
where $\Theta(\eta)$ is an arbitrary phase factor which may depend, in general, on $\eta$ .

 On the other hand, the matrix element $\bigr<\eta;t|\hat{\gamma}|\psi\bigr>$ of the operator $\hat{\gamma}$  between $\bigl<\eta;t|$ and $|\psi\bigr>$, where $|\psi\bigr>$ is an arbitrary state, is found from (35) and (112) to be 
\begin{equation}
\bigl<\eta;t|\hat{\gamma}|\psi\bigr>=\Theta^{\ast}(\eta)\int d q \bigl<\eta;t|q\bigr>\bigl<q|\hat{\gamma}|\psi\bigr>=\Bigl(\frac{\hbar}{i}\frac{\partial}{\partial \eta}-\frac{\hbar}{i\Theta^{\ast}}\frac{\partial \Theta^{\ast}}{\partial \eta}-\frac{\eta}{2\tilde{\varphi}(t)}\Bigr)\bigl<\eta;t|\psi\bigr>.
\end{equation}
Thus, if we take $\Theta(\eta)=\exp(\frac{i\eta^{2}}{4\hbar\tilde{\varphi}(t)})$, $\bigl<\eta;t| \hat{\eta}|\psi\bigr>$ and $\bigl<\eta;t|\hat{\gamma}|\psi\bigr>$ are given by
\begin{equation}
\bigl<\eta;t| \hat{\eta}|\psi\bigr>=\eta \bigl<\eta;t|\psi\bigr>,~~~ \bigl<\eta;t|\hat{\gamma}|\psi\bigr>=\frac{\hbar}{i}\frac{\partial}{\partial \eta}\bigl<\eta;t|\psi\bigr>,
\end{equation}
so that the $q$-representation of the eigenstate $|\eta;t\bigr>$, (112), turns out to be  
\begin{equation}
\bigl<q|\eta;t\bigr>=¡¡\bigl(4\pi \hbar f(t) \tilde{\varphi}(t)\bigr)^{-1/2}\exp\Bigl[\frac{i}{\hbar}\frac{f(t)\eta^{2}-2\eta q -2(g(t)\tilde{\varphi}(t)-\frac{1}{2f(t)})q^{2}}{4f(t)\tilde{\varphi}(t)}\Bigr].
\end{equation}

The $\eta$$-representation$ $\bigl<\eta;t|\psi \bigr>$ of an arbitrary state $|\psi \bigr>$ will then be obtained through  the $q$-representation $\bigl<q|\psi\bigr>$ of the same state as
\begin{equation}
\bigl<\eta;t|\psi \bigr>\equiv \int d q \bigl<\eta;t|q\bigr>\bigl<q|\psi \bigr>,
\end{equation} 
and  $\bigl<q|\eta;t\bigr>^{\ast}$, to be obtained from (115),  plays the role of the transformation function from the $q$- to the $\eta$-representations.

The $\eta$-representation $\bigl<\eta;t|\phi(t)\bigr>$ of the state $|\phi(t)\bigr>$ which is a solution of the Schr${\rm \ddot{o}}$dinger equation (58) is given by $\bigl<\eta;t|\phi(t)\bigr>\equiv \int d q\bigl<\eta;t|q\bigr>\bigl<q|\phi(t)\bigr>$, and is shown in fact to satisfy the differential equation
\begin{eqnarray}
i\hbar \frac{\partial}{\partial t}\bigl<\eta;t|\phi(t)\bigr>&=&i\hbar \frac{\partial}{\partial t}\int d q \bigl<\eta;t|q\bigr>\bigl<q|\phi(t)\bigr>  \nonumber  \\
&=&\int d q \Bigl[i\hbar \frac{\partial}{\partial t} \bigl<\eta;t|q\bigr> \cdot \bigl<q|\phi(t)\bigr>+\bigl<\eta;t|q\bigr>\cdot i\hbar \frac{\partial}{\partial t}\bigl<q|\phi(t)\bigr>\Bigr] =0,
\end{eqnarray}
where we have taken account of the fact that $\bigl<q|\phi(t)\bigr>$ satisfies the Schr${\rm \ddot{o}}$dinger equation
\begin{equation}
i\hbar \frac{\partial}{\partial t}\bigl<q|\phi(t)\bigr>=\Bigl[-\hbar^{2}X(t)\frac{\partial ^{2}}{\partial q^{2}}-i\hbar Y(t)\bigl(q\frac{\partial}{\partial q}+\frac{\partial}{\partial q}q\Bigr)+Z(t)q^{2}\Bigr]\bigl<q|\phi(t)\bigr>.
\end{equation} 
Eq. (117) means that in the HJ picture  the state does not evolve in the course of time. It is also possible to show directly  that in terms of a unitary operator $T(t)$ such as $|\eta;t\bigr>=T|\eta\bigr>$ the Hamiltonian $K$, to appear in the Schr$\ddot{{\rm o}}$dinger equation for $\bigl<\eta;t|\phi(t)\bigr>$, becomes $K=T^{-1}HT-i\hbar T^{-1}(\partial/\partial t)T=0$. Summarizing, our HJ picture corresponds, so to speak, to `looking at a moving body from the body- fixed moving reference frame'.

Lastly, we remark that since the operator $\hat{\eta}$ is defined here as a linear combination of $\hat{p}$ and $\hat{q}$, the above transformation from the $q$- to the $\eta$-representations is neither a point  transformation  nor the transformation from the $q$- to the $p$-representations. In this sense our $\eta$-representation is  of a new kind .

\subsection{Formal solutions in the HJ picture}
In order to obtain the $\eta$-representation of the eigenstate $|n;t\bigr>$ of the operator $\hat{a}(t)$, it is easier to proceed as follows (than via (116)). We first calculate  the matrix element  $\bigl<\eta;t|\hat{a}(t)|\psi\bigr>$  with respect to  an arbitrary state $|\psi\bigr>$;
\begin{eqnarray}
&&\bigl<\eta;t| \hat{a}(t)|\psi\bigr>=\int d q \bigl<\eta;t|q\bigr>\bigl<q|\hat{a}(t)|\psi\bigr>  \nonumber  \\
&&~~~=\frac{\Bigl[\frac{1}{2}\{2\sqrt{\kappa}+i(k_{3}-(k_{1}-k_{2})\tilde{\varphi}(t))\}\eta+\hbar\{k_{1}+k_{2}-k_{3}\tilde{\varphi}(t)-2i \sqrt{\kappa}\tilde{\varphi}(t)\}\frac{\partial}{\partial \eta}\Bigr]\bigl<\eta;t|\psi\bigr>}{\sqrt{2\hbar\sqrt{\kappa}\{k_{1}+k_{2}-2k_{3}\tilde{\varphi}(t )+(k_{1}-k_{2})\tilde{\varphi}^{2}(t )\}}}. \nonumber  \\
&&
\end{eqnarray}
Similarly we have
\begin{eqnarray}
&&\bigl<\eta;t| a^{\dagger}(t)|\psi\bigr>=\int d q \bigl<\eta;t|q\bigr>\bigl<q|a^{\dagger}(t)|\psi\bigr>  \nonumber  \\
&&~~~=\frac{\Bigl[\frac{1}{2}\{2\sqrt{\kappa}-i(k_{3}-(k_{1}-k_{2})\tilde{\varphi}(t))\}\eta-\hbar\{k_{1}+k_{2}-k_{3}\tilde{\varphi}(t)+2i \sqrt{\kappa}\tilde{\varphi}(t)\}\frac{\partial}{\partial \eta}\Bigr]\bigl<\eta;t|\psi\bigr>}{\sqrt{2\hbar\sqrt{\kappa}\{k_{1}+k_{2}-2k_{3}\tilde{\varphi}(t )+(k_{1}-k_{2})\tilde{\varphi}^{2}(t )\}}}, \nonumber  \\
&&
\end{eqnarray}
which can be rewritten as
\begin{eqnarray}
&=&-\sqrt{\frac{(k_{1}+k_{2})\hbar}{2\sqrt{\kappa}}}\Bigl(\frac{k_{1}+k_{2}-k_{3}\tilde{\varphi}(t)+2i\sqrt{\kappa}\tilde{\varphi}(t)}{k_{1}+k_{2}-k_{3}\tilde{\varphi}(t)-2i\sqrt{\kappa}\tilde{\varphi}(t)}\Bigr)^{1/2}  \nonumber  \\ 
&& \hspace{1cm}\times \exp\Bigl[\frac{2\sqrt{\kappa}-ik_{3}}{4\hbar(k_{1}+k_{2})}\eta^{2}\Bigr]\frac{\partial}{\partial \eta}\Bigl\{\exp \Bigl[-\frac{2\sqrt{\kappa}-ik_{3}}{4\hbar(k_{1}+k_{2})}\eta^{2}\Bigr]\Bigl<\eta;t|\psi\bigr>\Bigr\}.  \end{eqnarray}
Thus we have 
\begin{eqnarray}
\bigl<\eta;t|n;t\bigr>&=&\frac{1}{\sqrt{n!}}\bigl<\eta;t|\bigl( a^{\dagger}(t)\bigr)^{n}|0;t\bigr>  \nonumber  \\
&=&\frac{(-1)^{n}}{\sqrt{n!}}\Bigl(\frac{(k_{1}+k_{2})\hbar}{2\sqrt{\kappa}}\Bigr)^{n/2}\Bigl(\frac{k_{1}+k_{2}-k_{3}\tilde{\varphi}(t)+2i\sqrt{\kappa}\tilde{\varphi}(t)}{k_{1}+k_{2}-k_{3}\tilde{\varphi}(t)-2i\sqrt{\kappa}\tilde{\varphi}(t)}\Bigr)^{n/2}  \nonumber  \\
&&~~~~\times \exp\Bigl[\frac{2\sqrt{\kappa}-ik_{3}}{4\hbar(k_{1}+k_{2})}\eta^{2}\Bigr]\frac{\partial^{n}}{\partial \eta^{n}}\Bigl\{\exp\Bigl[-\frac{2\sqrt{\kappa}-ik_{3}}{4\hbar(k_{1}+k_{2})}\eta^{2}\Bigr]\bigl<\eta;t|0;t\bigr>\Bigr\}.
\end{eqnarray}
Further,  $\bigl<\eta;t|0;t\bigr>$  is found from
\begin{eqnarray}
\bigl<\eta;t|0;t\bigr>&=&\int d q \bigl<\eta;t|q\bigr>\bigl<q|0;t\bigr>  \nonumber  \\
&=&(-i)^{1/2}\Bigl(\frac{\sqrt{\kappa}}{\pi\hbar(k_{1}+k_{2})}\Bigr)^{1/4}\Bigl(\frac{k_{1}+k_{2}-k_{3}\tilde{\varphi}(t)+2i\sqrt{\kappa}\tilde{\varphi}(t)}{k_{1}+k_{2}-k_{3}\tilde{\varphi}(t)-2i\sqrt{\kappa}\tilde{\varphi}(t)}\Bigr)^{1/4}\exp\Bigl[-\frac{2\sqrt{\kappa}+ik_{3}}{4\hbar(k_{1}+k_{2})}\eta^{2}\Bigr] . \nonumber  \\
&&
\end{eqnarray}
Then, from (122) and (123) we have
\begin{eqnarray}
\bigl<\eta;t|n;t\bigr>& =&\frac{1}{\sqrt{n!}}\Bigl(\frac{-\sqrt{\kappa}}{\pi\hbar(k_{1}+k_{2})}\Bigr)^{1/4}\Bigl(\frac{k_{1}+k_{2}-k_{3}\tilde{\varphi}(t)+2i\sqrt{\kappa}\tilde{\varphi}(t)}{k_{1}+k_{2}-k_{3}\tilde{\varphi}(t)-2i\sqrt{\kappa}\tilde{\varphi}(t)}\Bigr)^{(n+\frac{1}{2})/2}  \nonumber  \\
&&~~~~~~~~\times \exp\Bigl[-\frac{2\sqrt{\kappa}+ik_{3}}{4\hbar(k_{1}+k_{2})}\eta^{2}\Bigr]H_{n}\Bigl(\sqrt{\frac{2\sqrt{\kappa}}{\hbar (k_{1}+k_{2})}}~\eta\Bigr),
\end{eqnarray}
which is the $\eta$-representation of the eigenstate $|n;t\bigr>$ of $\Lambda(t)$.

On the other hand, from (85), (53) and  (54)   we have
\begin{equation}
\exp\Bigl\{\frac{i}{\hbar}\int_{0}^{t} \theta_{n}(t^{\prime}) d t^{\prime}\bigr\}=\exp\Bigl(-i(n+\frac{1}{2})\varphi(t)\Bigr)=\Bigl(\frac{k_{1}+k_{2}-k_{3}\tilde{\varphi}(t)+2i\sqrt{\kappa}\tilde{\varphi}(t)}{k_{1}+k_{2}-k_{3}\tilde{\varphi}(t)-2i\sqrt{\kappa}\tilde{\varphi}(t)}\Bigr)^{- (n+\frac{1}{2})/2}.
\end{equation} 
Using (65), (124) and (125) , we find the $\eta$-representation $\bigl<\eta;t|\phi(t)\bigr>$ of the formal solution $|\phi(t)\bigr>$ as follows:
\begin{eqnarray}
\bigl<\eta;t|\phi(t)\bigr>&=&\sum_{n=0}c_{n}\exp\Bigl[\frac{i}{\hbar}\int_{0}^{t} \theta(t^{\prime})d t^{\prime}\bigr]\bigl<\eta;t|n;t \bigr>  \nonumber  \\
&=&\sum_{n}\frac{c_{n}}{\sqrt{n!}}\Bigl(\frac{-\sqrt{\kappa}}{\pi\hbar(k_{1}+k_{2})}\Bigr)^{1/4}\exp\Bigl[-\frac{2\sqrt{\kappa}+ik_{3}}{4\hbar(k_{1}+k_{2})}\eta^{2}\Bigr]H_{n}\Bigl(\sqrt{\frac{2\sqrt{\kappa}}{\hbar (k_{1}+k_{2})}}~\eta\Bigr).
\end{eqnarray}
As we   see in (124) and (126), the $\eta$-representation $\bigl<\eta;t|n;t\bigr>$ of the eigenstate $|n;t\bigr>$ explicitly depends on $t$, whereas  the $\eta$-representation $\bigl<\eta;t|\phi(t)\bigr>$ of the solution $|\phi(t)\bigr>$ of the Schr${\rm \ddot{o}}$dinger equation does not depend on $t$, being in agreement with (117). 

\subsection{The Feynman kernel in the HJ picture }
The Feynman kernel in the $\eta$-representation $\bigl<\eta_{1};t|U(t,0)|\eta_{2};0\bigr>$ is given, by use of (87), as
\begin{equation}
\bigl<\eta_{1};t|U(t,0)|\eta_{2};0\bigr>=\sum_{n}\bigl<\eta_{1};t|n;t\bigr>\exp\Bigl[\frac{i}{\hbar}\int_{0}^{t} \theta_{n}(t^{\prime}) d t^{\prime}\Bigr]\bigl<n;0|\eta_{2};0\bigr>,
\end{equation}
which is further  rewritten as 
\begin{eqnarray}
\bigl<\eta_{1};t|U(t,0)|\eta_{2};0\bigr>&=&\Bigl(\frac{\sqrt{\kappa}}{\pi \hbar (k_{1}+k_{2})}\Bigr)^{1/2} \exp\Bigl[-\frac{\sqrt{\kappa}(\eta_{1}^{2}+\eta_{2}^{2})}{2\hbar(k_{1}+k_{2})}\Bigr]\exp\Bigl[-\frac{ik_{3}(\eta_{1}^{2}-\eta_{2}^{2})}{4\hbar(k_{1}+k_{2})}\Bigr]  \nonumber  \\
&& ~~~\times \sum_{n}\frac{1}{n!}H_{n}\Bigl(\sqrt{\frac{2\sqrt{\kappa}}{\hbar(k_{1}+k_{2})}}~\eta_{1}\Bigr)H_{n}\Bigl(\sqrt{\frac{2\sqrt{\kappa}}{\hbar(k_{1}+k_{2})}}~\eta_{2}\Bigr),  \nonumber  \\
&&=\exp\Bigl[-\frac{i}{4\hbar}\frac{k_{3}(\eta_{1}^{2}-\eta_{2}^{2})}{k_{1}+k_{2}}\Bigr]\delta(\eta_{1}-\eta_{2})=\delta(\eta_{1}-\eta_{2}),
\end{eqnarray}
where we have used the formulae
\begin{equation}
\lim_{\beta\to 0} \sum_{n=0} \frac{e^{-\beta(n+1/2)}}{n!}H_{n}(\xi_{1}x_{1})H_{n}(\xi_{2}x_{2}) \exp\Bigl[-\frac{\xi_{1}^{2}x_{1}^{2}+\xi_{2}^{2}x_{2}^{2}}{4}\Bigr]=\sqrt{\frac{\pi}{2}}\delta\Bigl(\frac{\xi_{1}}{2}x_{1}-\frac{\xi_{2}}{2}x_{2}\Bigr).
\end{equation}
 
\subsection{Coherent states in the HJ picture}
The $\eta$-representation $\bigl<\eta;t|\phi(t)\bigr>_{\alpha}$ of the coherent state given by (93) becomes 
\begin{eqnarray}
\bigl<\eta;t|\phi(t)\bigr>_{\alpha}&=&\exp\Bigl[-\frac{|\alpha|^{2}}{2}-\frac{i}{2}\varphi(t) \Bigr]\sum_{n=0}^{\infty} \frac{(\tilde{\alpha}(t))^{n}}{\sqrt{n!}}\bigl<\eta;t|n;t\bigr>   \nonumber  \\
&=&e^{-|\alpha|^{2}/2}\Bigl(\frac{-\sqrt{\kappa}}{\pi\hbar(
k_{1}+k_{2})}\Bigr)^{1/4}\exp\Bigl[-\frac{2\sqrt{\kappa}+ik_{3}}{4\hbar(k_{1}+k_{2})}\eta^{2}\Bigr] \nonumber  \\
&& \hspace{2cm} \times \sum_{n=0}^{\infty}\frac{\tilde{\alpha}^{n}(t)}{n!}e^{i(n+1/2)\varphi(t)}H_{n}\Bigl(\sqrt{\frac{2\sqrt{\kappa}}{\hbar(k_{1}+k_{2}}}\eta^{2}\Bigr) \nonumber  \\
&=&e^{-\frac{|\alpha|^{2}}{2}}\Bigl(\frac{-\sqrt{\kappa}}{\pi\hbar(k_{1}+k_{2})}\Bigr)^{1/4}\exp\Bigl[-\frac{2\sqrt{\kappa}+ik_{3}}{4\hbar(k_{1}+k_{2})}\eta^{2}\Bigr]\sum_{n=0}^{\infty}\frac{ \alpha^{n}(t)}{n!}H_{n}\Bigl(\sqrt{\frac{2\sqrt{\kappa}}{\hbar(k_{1}+k_{2}}}\eta^{2}\Bigr)  \nonumber  \\
&=&\Bigl(\frac{-\sqrt{\kappa}}{\pi\hbar(k_{1}+k_{2})}\Bigr)^{1/4}\exp\Bigl[-\frac{|\alpha|^{2}}{2}-\frac{\alpha^{2}}{2}-\frac{2\sqrt{\kappa}+ik_{3}}{4\hbar(k_{1}+k_{2})}\eta^{2}+\alpha\sqrt{\frac{2\sqrt{\kappa}}{\hbar(k_{1}+k_{2})}}\eta\Bigr],
\end{eqnarray}
which no longer depends on $t$. We then have (by putting $\alpha=|\alpha|e^{-i\epsilon}$)
\begin{equation}
|\bigl<\eta;t|\phi(t)\bigr>_{\alpha}|^{2}=\Bigl(\frac{\sqrt{\kappa}}{\pi\hbar(k_{1}+k_{2})}\Bigr)^{1/2}\exp\Bigl[-\frac{\sqrt{\kappa}}{\hbar(k_{1}+k_{2})}\Bigl\{\eta-|\alpha| \sqrt{\frac{2\hbar(k_{1}+k_{2})}{\sqrt{\kappa}}}\cos \epsilon \Bigr\}^{2}\Bigr].
\end{equation}
From (131) it is seen that the classical limit of $|\bigl<\eta;t|\phi(t)\bigr>_{\alpha}|^{2}$ is 
\begin{equation}
\lim_{\hbar\to 0}|\bigl<\eta;t|\phi(t)\bigr>_{\alpha}|^{2}=\delta(\eta-\eta_{0}),
\end{equation}
provided that $|\alpha|$ is taken to be
\begin{equation}
|\alpha|=\sqrt{\frac{\sqrt{\kappa}}{\pi\hbar(k_{1}+k_{2})}}\frac{\eta_{0}}{\cos \epsilon}.
\end{equation}
This result shows that in the classical limit the particle moves on the trajectory $\eta=\eta_{0} ~(={\rm const.})$, in accordance with the classical situation where  $\eta$ is a constant of motion.

\section{Examples}
\subsection{A simple harmonic oscillator}
For the case where all coefficients  $X(t), Y(t)$ and  $Z(t)$ are  constants and equal, respectively, to 
\begin{equation}
X(t)=\frac{1}{2m},~~~Y(t)=0,~~~Z(t)=\frac{m\omega^{2}}{2},
\end{equation}
our system reduces to that of a simple harmonic oscillator. Here  $f(t)=\sqrt{\frac{1}{2m\omega}}\cos \omega t$ satisfies (42). Then  $f_{1}(t), f_{2}(t)$ in (50) and $\tilde{\varphi}(t)$ become 
\begin{equation}
f_{1}(t)=\sqrt{\frac{1}{2m\omega}}\cos \omega t,~~~~f_{2}(t)=-\sqrt{\frac{1}{2m\omega}}\sin \omega t,~~~~~\tilde{\varphi}(t)=\tan \omega t,
\end{equation}
which satisfy the relation $(\dot{f}_{1}f_{2}-f_{1}\dot{f}_{2})=X$.

The nonlinear differential equation (19) for $A(t)$ now takes the form:
\begin{equation}
\ddot{A}-\frac{\dot{A}^{2}}{2A}+2\omega^{2} A-\frac{2\kappa}{m^{2}}\frac{1}{A}=0.
\end{equation}
From (50) and (135) the solution $A(t)$ of (136) and the corresponding $B(t)$ and $C(t)$ are given by 
\begin{equation}
\begin{array}{l}
A=\frac{1}{2m\omega}(k_{1}+k_{2}\cos 2\omega t-k_{3}\sin 2\omega t),  \\
B=\frac{1}{2}(k_{2}\sin 2\omega t+k_{3}\cos 2\omega t),  \\
C=\frac{m\omega}{2}(k_{1}-k_{2}\cos 2\omega t+k_{3}\sin 2\omega t),
\end{array}
\end{equation}
where $(k_{1}^{2}-k_{2}^{2}-k_{3}^{2})=4\kappa$. Substituting (137) into (4),  $P$ is found to be  
\begin{eqnarray}
P&=&\frac{1}{2m\omega}(k_{1}+k_{2}\cos 2\omega t-k_{3}\sin 2\omega t)p^{2}+ (k_{2} \sin 2\omega t+k_{3}\cos 2\omega t)pq  \nonumber  \\
&&\hspace{1cm} +\frac{m\omega}{2}(k_{1}-k_{2}\cos 2\omega t+k_{3}\sin 2\omega t)q^{2}.
\end{eqnarray}
This is a constant of motion for a simple harmonic oscillator,  which explicitly depends on $t$  so far as   $k_{2} \ne 0$ and/or $k_{3} \ne 0$.

The Hamiltonian of a simple harmonic oscillator is a conserved quantity which does not depend on $t$. Our result shows  that the system has other constants of motion, which are  bilinear in  $p$ and $q$, and depend  on $t$  explicitly. Needless to say, $P(t)$ reduces to  $P=(2\sqrt{\kappa}/\omega)H$,  when $k_{2}=k_{3}=0 ~(k_{1}=2\sqrt{\kappa}$). In the following, we shall discuss the classical and quantum behavior of the simple harmonic oscillator by using $P$ given by (138).

\subsubsection{classical case  }
From (55) and (135)  we have
\begin{equation}
q=\sqrt{\frac{P}{2m \omega \kappa(k_{1}+k_{2})}}\Bigl\{(2\sqrt{\kappa}\cos \varphi_{0} -k_{3}\sin \varphi_{0}) \sin \omega t+ (k_{1}+k_{2})\sin \varphi_{0}\cos \omega t\Bigr\}.
\end{equation}
When $k_{2}=k_{3}=0 ~(k_{1}=2\sqrt{\kappa}$),  (139) reduces to the simple form
\begin{equation}
q=\sqrt{\frac{P}{m \omega \sqrt{\kappa}}} \sin(\omega t+\varphi_{0}).
\end{equation}
\subsubsection{quantum case}
From (135) and (137) we have 
\begin{eqnarray}
&& \frac{\xi(t)\xi(0)}{\sin \varphi(t)}=\frac{2m\omega}{\hbar \sin \omega t}, \hspace{1cm}~~ \frac{\cos \varphi(t)}{\sin \varphi(t)}=\frac{(k_{1}-k_{2})\cos \omega t+k_{3} \sin \omega t}{2\sqrt{\kappa}\sin \omega t},  \nonumber \\
&&-\frac{1}{2\hbar}\frac{B(t)}{A(t)}+\frac{\xi^{2}(t)\cos \varphi(t)}{4\sin \varphi(t)}=\frac{m\omega}{2\hbar}\frac{\cos \omega t}{\sin \omega t},  \\
&& \frac{1}{2\hbar}\frac{B(0)}{A(0)}+\frac{\xi^{2}(0)\cos \varphi(t)}{4\sin \varphi(t)}=\frac{m\omega}{2\hbar}\frac{\cos \omega t}{\sin \omega t}. \nonumber
\end{eqnarray}
Substituting these results into (88), we have 
\begin{equation}
\bigl<q_{1}|U(t,0)|q_{2} \bigr>=\Bigl(\frac{m\omega}{2\pi i\hbar \sin \omega t}\Bigr)^{1/2} \exp\Bigl[\frac{i m\omega }{2\hbar}\frac{\Bigl((q_{1}^{2}+q_{2}^{2})\cos \omega t -2q_{1}q_{2}\Bigr)}{\sin \omega t}\Bigr],
\end{equation}
which is nothing but the Feynman kernel of a simple harmonic oscillator . We note that the representation (142) of the kernel does not depend on the choice of $k_{1}, k_{2}$ and $k_{3}$, as required.

The dynamical phase and the geometrical phase are obtained from
\begin{eqnarray}
&&\bigl<n;t|H|n;t\bigr>=(n+\frac{1}{2})\hbar \omega ~\frac{ k_{1}}{2\sqrt{\kappa}},  \\
&&\bigl<n;t|i\hbar \frac{\partial}{\partial t}|n;t\bigr>=(n+\frac{1}{2})\hbar \omega ~\frac{k_{2}(k_{2}+k_{1}\cos 2\omega t )+k_{3}(k_{3}+k_{1} \sin 2\omega t)}{2\sqrt{\kappa}(k_{1}+k_{2} \cos 2\omega t-k_{3} \sin 2\omega t)},
\end{eqnarray}
and the uncertainty relation of the coherent state becomes 
\begin{equation}
(\triangle q)^{2}(\triangle p)^{2}=\frac{\hbar^{2}}{4}\Bigl(1+\frac{(k_{2}\sin 2\omega t+k_{3}\cos 2\omega t)^{2}}{4\kappa}\Bigr).
\end{equation} 
It is interesting that these expressions depend on   $k_{1}, k_{2}$ and  $k_{3}$. Particularly interesting is the fact  that for $k_{2}\ne 0$ and/or  $k_{3} \ne 0$ the geometrical phase does not vanish, and   the uncertainty relation does not have the minimum value . 

For $k_{2}=k_{3}=0 ~(k_{1}=2\sqrt{\kappa})$, on the other hand, (143), (144) and (145) reduce, respectively, to
\begin{equation}
\Bigl<n;t|H|n;t\bigr>=(n+\frac{1}{2})\hbar \omega,~~\bigl<n;t|i\hbar \frac{\partial}{\partial t}|n;t\bigr>=0,~~~(\triangle q)^{2}(\triangle p)^{2}=\frac{\hbar^{2}}{4},
\end{equation}
which are well-known results for a simple harmonic oscillator.

The above result indicates that the geometrical phase is not necessarily the quantity that is characteristic only  of the systems with $t$-dependent Hamiltonians . Evidently for the special case when $H$ does not depend on $t$,  we can choose the complete set of the ket vectors $|n;t\bigr>$ whose  geometrical phases are all vanishing.

\subsection{A damped oscillator}
The Hamiltonian $H$ of a damped oscillator depends on $t$ explicitly:  
\begin{equation}
H=\frac{1}{2m}e^{-2\mu t} p^{2}+\frac{m\omega^{2}}{2} e^{2\mu t} q^{2},
\end{equation}
and the equation of motion reads:
\begin{equation}
\ddot{q}+2\mu \dot{q}+\omega^{2} q=0,
\end{equation}
where $\mu$ is an arbitrary constant satisfying the relation $\mu <\omega$.
In this system, $X(t), Y(t)$ and $Z(t)$ are given by
\begin{equation}
X(t)=\frac{1}{2m}e^{-2\mu t}, \hspace{1cm} Y(t)=0, \hspace{1cm} Z(t)= \frac{m\omega^{2}}{2}e^{2\mu t},
\end{equation}
so that (19) becomes 
\begin{equation}
\ddot{A}-\frac{\dot{A}^{2}}{2A}+2\mu \dot{A}+2\omega^{2} A-\frac{2\kappa}{m^{2}}\frac{e^{-4\mu t}}{A}=0.
\end{equation}

Since in this case, $f_{1}(t), f_{2}(t)$ and $\tilde{\varphi}(t)$ are given by
\begin{equation}
f_{1}(t)=\sqrt{\frac{1}{2m\tilde{\omega}}}e^{-\mu t}\cos \tilde{\omega}t,~~f_{2}(t)=-\sqrt{\frac{1}{2m\tilde{\omega}}}e^{-\mu t}\sin \tilde{\omega}t,~~\tilde{\varphi}(t)=\tan \tilde{\omega}t,
\end{equation}
where $\tilde{\omega}\equiv \sqrt{\omega^{2}-\mu^{2}}$, we can proceed in the same way as in the previous example, the result being
\begin{eqnarray}
A(t)&=&\frac{e^{-2\mu t}}{2m\tilde{\omega}}(k_{1}+k_{2} \cos 2\tilde{\omega}t -k_{3}\sin 2\tilde{\omega}t),  \\
B(t)&=&\frac{1}{2}(k_{2} \sin 2\tilde{\omega}t+k_{3}\cos 2\tilde{\omega} t) +\frac{\mu}{2\tilde{\omega}}(k_{1}+k_{2} \cos 2\tilde{\omega}t -k_{3}\sin 2\tilde{\omega}t),  \\
C(t)&=&\frac{m\tilde{\omega}}{2}e^{2\mu t}(k_{1}-k_{2} \cos 2\tilde{\omega}t +k_{3}\sin 2\tilde{\omega}t) +m\mu e^{2\mu t}(k_{2} \sin 2\tilde{\omega}t +k_{3}\cos 2\tilde{\omega}t)  \nonumber  \\
&&+\frac{m \mu^{2}}{2\tilde{\omega}}e^{2\mu t}(k_{1}+k_{2} \cos 2\tilde{\omega}t -k_{3}\sin 2\tilde{\omega}t).
\end{eqnarray}
where  $(k_{1}^{2}-k_{2}^{2}-k_{3}^{2})=4\kappa$.

Then the conserved quantity $P$ defined by (4) takes the form
\begin{eqnarray}
P&=&(k_{1}+k_{2} \cos 2\tilde{\omega}t -k_{3}\sin 2\tilde{\omega}t)\Bigl(\frac{e^{-2\mu t}}{2m\tilde{\omega}}p^{2}+\frac{\mu}{\tilde{\omega}}pq+\frac{m\mu^{2}}{2\tilde{\omega}}e^{2\mu t} q^{2}\Bigr)  \nonumber  \\
&&+(k_{2} \sin 2\tilde{\omega}t +k_{3}\cos 2\tilde{\omega}t) \bigl(pq+m\mu e^{2\mu t}q^{2}\bigr)  \nonumber  \\
&&+\frac{m\tilde{\omega}}{2}e^{2\mu t}(k_{1}-k_{2} \cos 2\tilde{\omega}t +k_{3}\sin 2\tilde{\omega}t)q^{2}.
\end{eqnarray}
When $k_{2}=k_{3}=0~(k_{1}=2\sqrt{\kappa})$, the above $P$ turns out to be
\begin{equation}
P=\frac{2\sqrt{\kappa}}{\tilde{\omega}}(H+\mu pq).
\end{equation}

\subsubsection{classical case}
The classical solution $q$ of this system is given by
\begin{equation}
q=\sqrt{\frac{P}{2m \tilde{\omega} \kappa(k_{1}+k_{2})}} e^{-\mu t} \Bigl\{(2\sqrt{\kappa}\cos \varphi_{0} -k_{3}\sin \varphi_{0}) \sin \tilde{\omega} t+ (k_{1}+k_{2})\sin \varphi_{0}\cos \tilde{\omega} t\Bigr\}.
\end{equation}
When $k_{2}=k_{3}=0  ~(k_{1}=2\sqrt{\kappa}$), (157) reduces to
\begin{equation}
q=\sqrt{\frac{P}{m \tilde{\omega} \sqrt{\kappa}}}e^{-\mu t} \sin(\tilde{\omega} t+\varphi_{0}).
\end{equation}
\subsubsection{quantum case}
In a way similar to the previous example, we can obtain the Feynman kernel:   
\begin{eqnarray}
\bigl<q_{1} |U(t,0)|q_{2}\bigr>&=&\Bigl(\frac{m\tilde{\omega}}{2\pi i \hbar \sin \tilde{\omega}t}\Bigr)^{1/2}e^{\mu t/2}  \nonumber  \\
&& \times \exp\Bigl[ \frac{i m\tilde{\omega}}{2\hbar}e^{2\mu t}(-\mu+\frac{\cos\tilde{\omega}t}{\sin \tilde{\omega}t})q_{1}^{2}+\frac{i m\tilde{\omega}}{2\hbar}(\mu+\frac{\cos\tilde{\omega}t}{\sin \tilde{\omega}t})q_{2}^{2} -\frac{i m \tilde{\omega}}{\hbar}\frac{e^{\mu t}}{\sin \tilde{\omega}t} q_{1}q_{2}\Bigr]. \nonumber  \\
&&
\end{eqnarray}
The dynamical   and  geometrical phases are then found from
\begin{eqnarray}
\bigl<n;t|H|n;t\bigr>&=&(n+1/2)\hbar\omega (2\sqrt{\kappa}\omega\tilde{\omega})^{-1}\Bigl\{(k_{1}\omega^{2} +k_{2}\mu^{2} \cos 2\tilde{\omega}t-k_{3}\mu^{2} \sin 2\tilde{\omega}t)  \nonumber  \\
&&\hspace{5cm}+\frac{\mu}{2}(k_{2} \sin 2 \tilde{\omega}t+k_{3}\sin 2 \tilde{\omega}t )\Bigr\},  \\
\bigl<n;t|i \hbar \frac{\partial}{\partial t}|n;t\Bigr>&=& (n+\frac{1}{2})\hbar \omega (2\omega \tilde{\omega}\sqrt{\kappa})^{-1}(k_{1}+k_{2}\cos 2\tilde{\omega}t -k_{3} \sin 2\tilde{\omega}t)^{-1}  \nonumber  \\
&&\times [(\mu^{2}k_{1}^{2}+\omega^{2}k_{2}^{2}+\omega^{2}k_{3}^{2} )+k_{1}(\omega^{2}k_{2}+\mu^{2}k_{2}+\mu \tilde{\omega}k_{3}) \cos 2\tilde{\omega}t  \nonumber  \\
&&\hspace{0.5cm}-k_{1}(\omega^{2}k_{3}+\mu^{2}k_{3}-\mu \tilde{\omega}k_{2}) \sin 2\tilde{\omega}t-\mu (\mu k_{2}^{2}+ \tilde{\omega}k_{2}k_{3})\sin^{2} 2\tilde{\omega}t  \nonumber  \\
&& \hspace{0.5cm}-\mu(\mu k_{3}^{2}-\tilde{\omega}k_{2}k_{3})\cos^{2}\tilde{\omega}t+\frac{1}{2}\mu(-2\mu k_{2}k_{3}+\tilde{\omega}k_{2}^{2}-\tilde{\omega}k_{3}^{2})   \sin 4\tilde{\omega}t ].  \nonumber  \\
&&
\end{eqnarray}
For the case $k_{2}=k_{3}=0 ~(k_{1}=2\sqrt{\kappa}$), (160) and (161) reduce to
\begin{equation}
\bigl<n;t|H(t)|n;t\bigr>=(n+\frac{1}{2})\hbar \omega ~\Bigl(\frac{\omega}{\tilde{\omega}}\Bigr),~~~\Bigl<n;t|i\hbar \frac{\partial}{\partial t}|n;t\bigr>=(n+\frac{1}{2})\hbar \omega ~\Bigl(\frac{\mu^{2}}{\omega \tilde{\omega}}\Bigr).
\end{equation}
It is to  be noticed that the geometrical phase   does not vanish so far as $\mu \ne 0$. 

\section{Final Remarks}

\noindent a) It is of interest to recall that both the classical and quantum mechanical behavior of generalized oscillators is basically determined by one and the same function $A(t)$ which satisfies a nonlinear differential equation.

\vspace{0.5cm}

\noindent b) Of interest also is a mathematical theorem such as exhibited by equation (50). Here, a solution $A(t)$ to the $ nonlinear$ differential equation (19) is given in terms of other functions $f_{1}(t)$ and $f_{2}(t)$, each being a solution to the $linear$ differential equation (42). In addition to the physical reason for this, given in Subsect. 2.3, there may perhaps exist some deeper mathematical reason, of  which   further investigation is hoped for.

\vspace{0.5cm}

\noindent c) Even for the case of a simple harmonic oscillator there are many ways of constructing coherent states, as seen from the results of Sect. 4, and the uncertainty in such states does not, in general, take the minimum value.

\vspace{0.5cm}

\noindent d) In the case when $X(t)=X_{0}+\triangle X(t), Y(t)=Y_{0}+\triangle Y(t)$ and $Z(t)=Z_{0}+\triangle Z(t)$;  $\triangle X(t)/X_{0} \ll 1,~ \triangle Y(t)/Y_{0} \ll 1$ and $\triangle Z(t)/Z_{0}\ll 1$ with  $X_{0}, Y_{0}$ and $Z_{0}$ being  constants, eq. (19) has the approximate solution   $A(t)=A_{0}+\triangle A(t)$ with $A_{0} \equiv \sqrt{\kappa/(X_{0}Z_{0}-Y_{0}^{2})}~X_{0}$ and with  $\triangle A(t)$ satisfying
\begin{equation}
\ddot{\triangle A(t)}=-4A_{0}\Bigl[2X_{0}Z_{0}\bigl\{\frac{\triangle Z(t)}{Z_{0}}-\frac{\triangle X(t)}{X_{0}}\bigr\}+4Y_{0}^{2}\bigl\{\frac{  \triangle X(t)}{X_{0}}-\frac{\triangle Y(t) }{Y_{0}}\bigr\}+Y_{0}\bigl\{\frac{  \dot{\triangle X(t)}}{X_{0}}- \frac{\dot{\triangle Y(t)} }{Y_{0}}\bigr\} \Bigr].
\end{equation}
The solution thereby obtained is good up to  the first order of $ \triangle X_{0}/X_{0},  \triangle Y_{0}/Y_{0}$ and $\triangle Z_{0}/Z_{0}$.  

\begin{acknowledgements}
We are indebted to  Professor Satoru Saito for useful discussions.
\end{acknowledgements}

\end{document}